\documentstyle[preprint,aps]{revtex}

\title{Evidence for a Pseudogap Above the Critical
Temperature in the Attractive Hubbard Model.}\\
\author{J.J.
Rodr\'{\i}guez-N\'u\~nez$^{1,3}$\footnote
{Permanent address: Grupo de F\'{\i}sica de S\'olidos,
Departamento
de F\'{\i}sica, FEC - LUZ, Apartado 526, Maracaibo,
{\bf Venezuela}}, S. Schafroth$^2$,
R. Micnas$^1$,\\
T. Schneider$^1$, H. Beck$^3$ and M.H. Pedersen$^{1,2}$.}
\address{$^1$IBM Research Division, Z\"urich
Research Laboratory,
Saumerstrasse 4, R\"uschlikon, {\bf Switzerland},
$^2$Physik-Institut der
Universit\"at Z\"urich, Winterthurerstrasse 190,
CH-8057 Z\"urich, {\bf Switzerland},
$^3$Institut de Physique,
Universit\'e de
Neuch\^atel,
Rue A.L. Breguet 1, CH-2000 Neuch\^atel,
{\bf Switzerland}.}
\begin{document}
\draft
\maketitle
\begin{abstract}
We explore the effect of
fluctuations for $T \geq T_c$ in
the 2-D negative Hubbard model within the
framework of the selfconsistent
T-matrix formalism, which
goes beyond the BCS approximation and
includes pair fluctuations.
We enter the regime where correlations
are important, namely, $U/t = -4.0$,
where $t$ denotes the hopping
matrix element between nearest neighboors and $U$
is the strength of the on-site interaction. Our results
include: 1) In comparison to the free case, the
distribution function, $n({\bf k})$, reveals
considerable renormalization. This fact makes us
conclude that the Fermi
surface is blurred for correlated
electron systems; 2) There is a pseudogap in the
density of states, $N(\omega)$, around the
chemical potential,
$\mu$; and 3) The real part of the T-matrix, at
zero frequency and zero momentum, vs $T/t$
shows a divergence at a
particular temperature. This last
result shows that more approximate T-matrix calculations,
extended to $T~=~0$ are meaningless.
\end{abstract}

\newpage

    High-$T_c$ superconducting materials
are extreme type-II
superconductors, with short correlation
length and pronounced
uniaxial anisotropy. As a consequence,
thermal fluctuations are
essential and known to mediate critical point
universality.\cite
{Schneider-Keller}. These fluctuations
invalidate mean-field
theories around $T_c$, but away from $T_c$,
the pronounced
uniaxial anisotropy of cuprates
leads to a crossover from three dimensions
(3D) to quasi 2D-behavior, where
fluctuations are essential,
again. In this Rapid Communication, we
explore the pair
fluctuation effects above $T_c$
in the 2-D attractive Hubbard
model\cite{Micnas et al} within the
framework of the fully self-consistent
T-matrix formalism\cite{Kadanof and Baym}
and in the regime where weak coupling treatments
do not apply.
Previous fully self-consistent
T-matrix calculations have been performed
numerically\cite{Fresard,Diploma} for a weaker coupling
than ours and analytically\cite{Haussmann} in both,
but for the special case of parabolic bands. The pairing
of electrons, which does not necessarily imply the
occurrence of superconductivity, is usually
described by the anomalous
Green's function, $F({\bf x}\tau,{\bf x'}\tau')$.
In terms of the operator\\
\begin{equation}
Q({\bf x}\tau)~\equiv~c_{\downarrow}({
\bf x}\tau)c_{\uparrow}({\bf x}\tau)~~,
\end{equation}
the expectation value\\
\begin{equation}
\Delta~\equiv~-U <Q({\bf x}\tau)>~~,
\end{equation}
\noindent
defines the pairing order parameter below $T_c$.\\

   Beyond BCS one has to include pair fluctuations above
$T_c$. For this purpose, we
introduce the pair Green's function,
$G_2$, defined as
follows\\
\begin{equation}\label{number}
G_2({\bf x}\tau,{\bf x'}\tau')~\equiv~<T_{\tau}
Q({\bf x}\tau)
Q^{\dagger}({\bf x'}\tau')>~~.
\end{equation}

    In the weak coupling limit (and
above $T_c$), the pair Green's
function factorizes\\
\begin{equation}\label{number2}
G_2({\bf x}\tau,{\bf x'}\tau')~
=~
G({\bf x}\tau,{\bf x'}\tau')
G({\bf x}\tau,{\bf x'}\tau')~~,
\end{equation}
\noindent
where, $T_{\tau}$ means time ordering.
For a finite $U$, however, there are additional
contributions, due to the correlations between
particles. In fact, the evaluation of Eq.
(\ref{number}) (in Fourier space) yields:\\
\begin{equation}\label{number3}
G_2({\bf q},i\epsilon_m)~=~\frac{\chi({\bf q},i\epsilon_m)
T({\bf q},i\epsilon_m)}{U}~~,
\end{equation}
\noindent
where $T$ is the T-matrix and
$\chi$ is given by the right hand side of
Eq. (\ref{number2}), which in $k$-space is
written as\\
\begin{equation}
\chi({\bf q},i\epsilon_m)~\equiv~
\frac{1}{\beta N}\sum_{{\bf k},\omega_n}G({\bf k},
i\omega_m)G({\bf q -  k},
i\epsilon_m - i\omega_n)~~,
\end{equation}
\noindent
where $\epsilon_m$($\omega_n$) are the bosonic
(fermionic)
Matsubara frequencies. The $T$-matrix is
expressed as\\
\begin{equation}\label{number4}
T({\bf q},i\epsilon_m)~\equiv~\frac{U}{1 + U\chi({\bf q},
i\epsilon_m)}~~.
\end{equation}

   The difference with respect to
the free case is that
the T-matrix  at
zero frequency and zero momentum diverges  at a
particular temperature. This temperature
can be taken as an estimate for the critical temperature,
$T_c$, indicating the instability of the normal
state against gauge symmetry
breaking (Thouless criterion).\\

   In this report we present
numerical results for $U/t~=~-4.0$, including:
1) the distribution function, $n({\bf k})$;
2) the density
of states, $N(\omega)$; 3) the phase shift, $\delta \phi
({\bf q},\omega)$; 4) the real
part of the T-matrix at zero frequency and at zero
momentum.  For this value of $U/t$,
correlations are important, since $|U|$ is half of the
bandwidth, $W$. ($W~=~8t$ in 2D).\\

   In Fig. 1 we show the distribution function,
$n({\bf k})$, along the diagonal
of the Brillouin zone, i.e.,
${\bf k} \equiv (Q,Q)$ and we compare
it with respect to the
free distribution function, i.e.,
for $U/t~=~0$. The temperature
chosen is $T/t~\equiv~0.125$. We have
calculated the parameter
$\alpha$\cite{Pines and Nozieres},
which measures the relative number
of excited quasiparticles.
With $32\times32$ points in the Brillouin
zone and 1024 Matsubara
frequencies, we get that
$\alpha~\equiv~20~\%$. This implies
that $10~\%$ of electrons are now excited
from below to above the Fermi wave vector of the
noninteracting system.
This would imply that the Fermi
surface is probably not a well
defined quantity, for this value
of the interaction, since
the distribution function has a less abrupt change as
function of momentum than for the free case.\\

    Another quantity of interest is the
density of states (DOS), $N(\omega)$.
In Fig. 2 we show the DOS for
$n/2~\equiv~0.10$ for different
temperatures. We
observe that, for a temperature
of $T/t~\equiv~0.22$, there
is no pseudogap. Then, for temperatures
between 0.20 and 0.17, we
observe the opening of the pseudogap. The
reason that the pseudogap does not
fully develop when the temperature
decreases is
probably due to the choice
of the damping factor,
$\delta~\equiv~0.25$. This
choice is made to avoid any additional structure which
could appear away from the chemical
potential due to finite
size effects.
We should mention that the
temperature of pair formation, $T_p$ can
be estimated in different ways. The first one
is the one given
by a mean field BCS approximation.
The second one is to evaluate numerically
the disappearance of the
pseudogap in the DOS.\cite{Flores}
This way of calculating
$T_p$ is probably not too precise.
We note that the presence of
fluctuations above $T_c$ can produce the opening of
a pseudogap in
DOS within the framework of the Ginzburg -
Landau approach in the
self-consistent Hartree approximation
(if a cutoff parameter
is extended down to the lattice scale)
by allowing randomness
in $\Delta({\bf r})$ down to the atomic
scale.\cite{Park and
Joynt} From a microscopic point of view, Moreo et
al\cite{Moreo},
by Quantum Monte Carlo simulations,
for $U/t~=~-4.0$ and $n~=~0.87$,
have found a pseudogap above $T_c$ and a
fully developed gap  at $T_c$.
Our formalism
is more suitable for lower densities.\\

    The existence of fluctuating pairs above the critical
temperature is
also signaled by the appearence of
$\Theta$-like functions in the
phase shift vs frequency, for large momentum.
(We refer to the
work of Refs.\cite{NSR,Varma et al}).
The phase shift
is defined as\\
\begin{equation}
\delta \phi({\bf q},\omega) \equiv
tan^{-1}\left( \frac{Im[T({\bf q},\omega)]}{Re[
T({\bf q},\omega)]}\right) ~~,
\end{equation}
\noindent
where Re(Im)(...) means real (imaginary) part of what
follows.\\

     From our Fig. 3 we observe that for
large momenta, there appears a
$\Theta$-like behavior which
approaches $\pi$. We
observe some features which resemble the appearence of
bound or resonant states.\\

   In a full self consistent
T-matrix formalism, we have the critical temperature,
which can be estimated in terms of the real
part of the T-matrix at zero frequency
and at zero momentun.\cite{Kadanof and Baym} At $T_c$,
this quantity diverges and signals the formation of
stable pairs,  i.e., a symmetry breaking and
$\Delta~\neq~0$. Below this
temperature, however, the formalism
must be generalized. (Some authors
\cite{Varma et al} have extended
the T-matrix approach
down to zero temperatures without
taking a finite $\Delta$. What
they get is that the chemical
potential, for any density and for any coupling
strength, approaches
that given by bound pairs. (See, for exemple,
Fig. 3 of Ref. \cite
{Varma et al}). In Fig. 4 we present the real part of the
T-matrix vs temperature
for $U/t~\equiv~-4.0$ and $n/2~\equiv~0.05.$,
revealing a divergence  around $T/t~=~0.16$. The critical
temperature is a quantity which has to be calculated by a
fully self-consistent T-matrix approach, i.e., a
perturbative calculation will give a different
value for $T_c$.\cite{spain} To check our
estimates for $T_c$, we consider the chemical
potential as function of temperature
for the same density and same number
of k-points and Matsubara frequencies.
The numerical data shown in Fig. 5,
even when we have artificially extended
our calculations below $T_c$. is very close to that
of Ref. \cite{Varma et al}. However,
we should say that we cannot use the
T-matrix formalism below the
critical temperature, unless we modify the
formalism to include off diagonal one particle-Green's
functions.\\

   In conclusion, we have evaluated several
quantities in the fully self-consistent
T-matrix formalism and in the
regime where correlations are importante for the
attractive Hubbard model.
In contrast to the BCS weak coupling approximation,
the presence of
fluctuating pairs above $T_c$ has important
effects on the density of states, the
phase shift and the distribution
function. This distribution
function indicates that the
Fermi surface is more blurred
in correlated systems than in weakly
interacting ones. In other
words, the full self-consistent
T-matrix approach can be used more
properly to describe the high-$T_c$ superconductors,
since they are characterized by strong fluctuations.
In addition,
we have used the fully T-matrix formalism to evaluate the
critical temperature. The latter result shows that
previous calculations which have used the not
conserving T-matrix approach
down to zero temperature is not
correct unless it is modified
to include anomalous Green's functions.
The presence of pairs
also has some dramatic effect on other electronic
quantities, like the spectral function
for the single
particle Green's function.
These results will be published
elsewhere. Finally, we should
remark that in the T-matrix formalism
the Thouless criterion
is formally equivalent to the
appearence of a pairing order parameter,
i.e., $\Delta~\neq~0$.\\

   We gratefully acknowledge the support of the Swiss
National Science Foundation. JJRN thanks partial
support from project
N$^o$. F-139 (CONICIT) and CONDES-LUZ.
R.M. acknowledges the partial support from the
Committee for Scientific Research (KBN Poland,
project No. 2 P3 02 057 04).
We would like to thank Ms.
Mar\'{\i}a Dolores Garc\'{\i}a-Gonz\'alez for reading
the manuscript. We have profited from interesting
discussions with T. MeintruJ.-P.
Locquet and C. Rossel.

\newpage
\vspace{0.8cm}
\begin{center}
{\bf Figure Captions.}\\
\end{center}
\vspace{1.0cm}
\noindent
Figure 1.- The distribution function, $n({\bf k})$, for
$U/t~=~-4.0$, $T/t~=~0.125$, $n/2~=~0.16$. We have chosen
a damping factor, $\delta$, in the Green function of
0.30 to go to real frequencies.
The number of lattice sites is $N_x\times
N_y~=~32\times 32$
and the number
of Matsubara frequencies is $NMats~=~1024$.
$n$ denotes the number of electrons per site.
We show only 11 points of the Brillouin zone and
the units are $\frac{\pi \times Q}{16}$\\
Figure 2.- The Density of States, $N(\omega)$ for
different temperatures. Here, we have chosen the same
interaction as in Fig. 1. The density per spin is
$n/2~=~0.10$ and the damping factor is $\delta~=
are the same as in Fig. 1.
The energy is measured
with respect to the chemical potential.\\
Figure 3.- The phase shift for
different values of momenta along
the diagonal of
the Brillouin zone. The same parameters
as in Fig. 2.\\
Figure 4.- Real part of the T-matrix at
zero frequency and
zero momentum vs. temperature. The interaction
is the same as in Fig. 1. The density per spin is
0.05 and the damping factor is 0.1.
$N_x\times N_y~=~22\times 22~$ and $NMats~=~1024$.\\
Fig. 5.- $T/t~$ vs. $~\mu/t$ for the
same parameters as in Fig. 4. The form presented here is
similar to the one given in Ref. \cite{Varma et al}.\\
\end{document}